\documentclass[]{IEEEphot}

\jvol{xx}
\jnum{xx}
\jmonth{June}
\pubyear{2009}

\usepackage{cite}
\begin{document}

\title{Computational Study of Amplitude-to-Phase \\Conversion in a Modified Uni-Traveling Carrier (MUTC) Photodetector}

\author{Yue Hu,$^1$~\IEEEmembership{Student~Member,~IEEE},
Curtis R. Menyuk,$^1$~\IEEEmembership{Fellow,~IEEE},  
Xiaojun  Xie,$^2$\\ 
Meredith N. Hutchinson,$^3$~\IEEEmembership{Senior~Member,~IEEE},
Vincent J. Urick,$^3$~\IEEEmembership{Senior~Member,~IEEE},\\
Joe C. Campbell,$^2$~\IEEEmembership{Fellow,~IEEE},
Keith J. Williams$^3$,~\IEEEmembership{Senior~Member,~IEEE}}

\affil{$^1$Computer Science and Electrical Engineering Department, University of Maryland Baltimore County, Baltimore, MD, 21250 USA\\
	$^2$Electrical and Computer Engineering, University of Virginia,Charlottesville, VA 22904, USA\\
	$^3$Naval Research Laboratory, Photonics Technology Branch, Washington, DC 20375, USA
	}  

\doiinfo{DOI: 10.1109/JPHOT.2016.XXXXXXX\\
1943-0655/\$25.00 \copyright 2016 IEEE}%

\maketitle

\markboth{IEEE Photonics Journal}{Volume Extreme Ultraviolet Holographic Imaging}

\begin{receivedinfo}%
Manuscript received March 3, 2008; revised November 10, 2008. First published December 10, 2008. Current version published February 25, 2009. This research was sponsored by the National Science Foundation through the NSF ERC Center for Extreme Ultraviolet Science and Technology, NSF Award No. EEC-0310717. This paper was presented in part at the National Science Foundation.
\end{receivedinfo}

\begin{abstract}
We calculate the amplitude-to-phase (AM-to-PM) noise conversion in a  modified uni-traveling carrier (MUTC) photodetector.  We obtained two nulls as measured in the experiments, and we explain their origin. The nulls appear due to the transit time variation when the average photocurrent varies, and the transit time variation is due to the change of electron velocity  when the average photocurrent varies. We also show that the AM-to-PM conversion coefficient depends only on the pulse energy and is independent of the pulse duration when the duration is less than 500 fs. When the pulse duration is larger than 500 fs, the nulls of the AM-to-PM conversion coefficient shift to larger average photocurrents. This shift occurs because the increase in that pulse duration leads to a decrease in the peak photocurrent. The AM-to-PM noise conversion coefficient changes as the repetition rate varies. However, the repetition rate does not change the AM-to-PM conversion coefficient as a function of input optical pulse energy. The repetition rate changes the average photocurrent. We propose a design that would in theory improve the performance of the device.
\end{abstract}


\section{Introduction}
Ultrastable microwave signals are of great interest in applications such as radar, telecommunications, navigation systems, and time synchronization \cite{kim2008drift,jung2015all,ghelfi2014fully}. Recently, there has been great interest in generating microwaves through optical frequency division (OFD) with a modelocked laser comb. Ultrastable microwave generation has been demonstrated \cite{Diddams:09,fortier2011,Fortier2013,Xie2015}. The stability of the optical reference is transferred to the repetition rate of  the pulse train of a modelocked laser. Two factors determine the noise level of the microwave output. One factor is the phase noise in the modelocked laser comb, and the other factor is the phase noise that is generated in the photodetector. In this paper, we will focus on the second factor, explaining in particular the amplitude-to-phase noise conversion in a high current photodetector. 

The phase noise that is produced by the photodetector is a critical limit to system performance. A major source of phase noise is amplitude-to-phase (AM-to-PM) conversion in the photodetector. Zhang et al.~\cite{Zhang2012} used a simple model to study the AM-to-PM conversion coefficient in a $p$-$i$-$n$ photodetector. Taylor et al.~\cite{Taylor2011} experimentally studied AM-to-PM conversion and found that there are several nulls, where the AM-to-PM conversion coefficient approaches zero as a function of the photocurrent.  In the experiments, they use the impulse response method to measure the phase change and the AM-to-PM noise conversion coefficient.

In this paper, we will use a modified drift-diffusion model to study the AM-to-PM conversion in the modified uni-traveling carrier (MUTC) photodetector. We first calculate the impulse response in the time domain and ,from that, we calculate the AM-to-PM conversion coefficient. We will explain the physical origin of the conversion nulls, and we will show how the pulse duration affects the AM-to-PM noise conversion coefficient. We will show that when the pulse duration is shorter than 500 fs, the output is only affected by the total pulse energy.

\section{Model}
In the paper, we use a drift-diffusion model \cite{Huyue2014,Hu:15} to study the AM-to-PM conversion in an MUTC photodetector \cite{Li2010,hu2016} with the configuration shown in Fig.~1.

The AM-to-PM conversion coefficient $\alpha$ is defined as the phase change in the device $\Delta\phi$, divided by the fractional optical power fluctuation $\Delta P/P$.
\begin{equation}
\alpha=\frac{\Delta\phi}{\Delta P/P}. \label{AMPM}
\end{equation}
\begin{figure}[!h]
	\centerline{
		{\includegraphics[width=12cm]{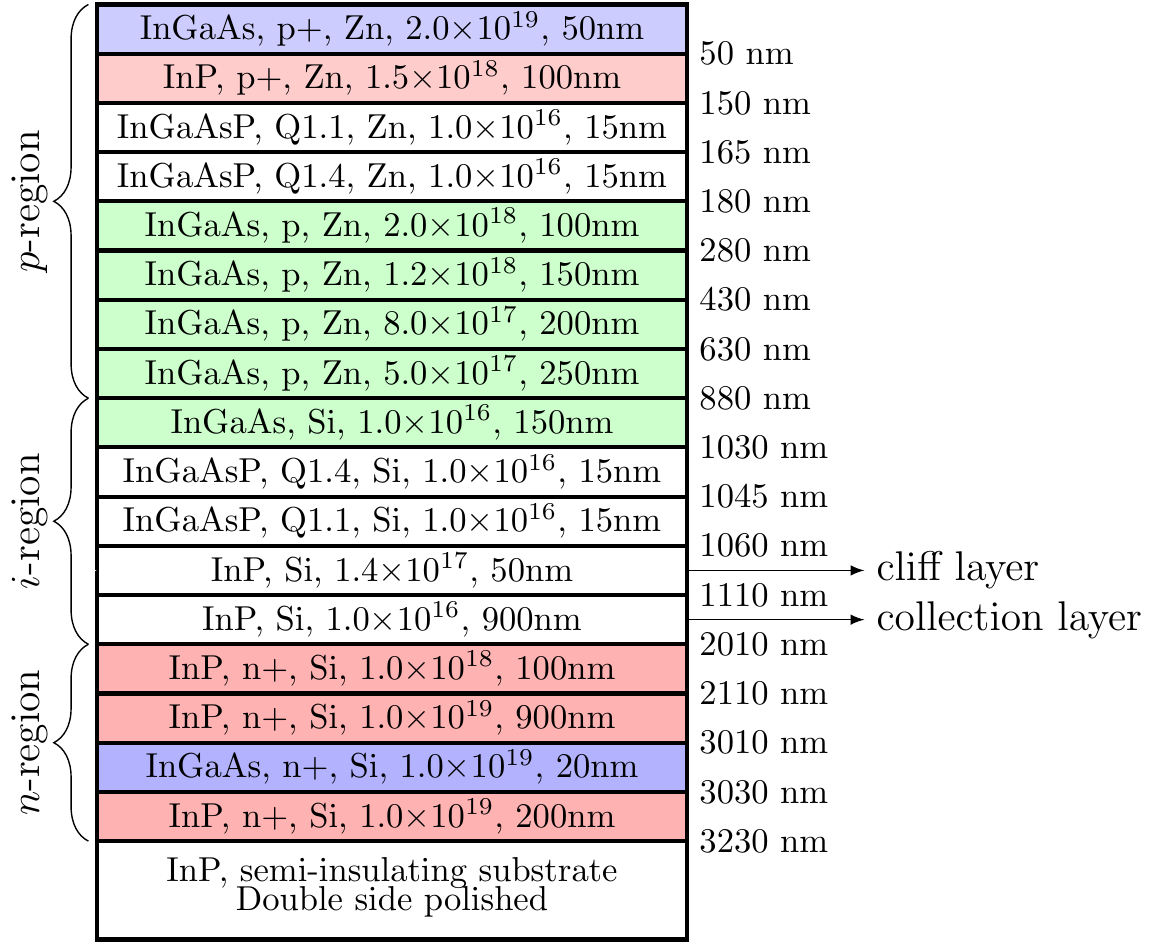}\label {Structure}}
	}
	\caption{Structure of the MUTC photodetector. Green indicates the absorption regions, which includes an intrinsic absorption region and a $p$-doped absorption region. Red indicates the highly-doped InP layers, and purple indicates highly-doped InGaAs, and white indicate other layers..}
\end{figure}

There are several methods to measure the phase noise \cite{Taylor2011}. One is to use a phase bridge, which directly measures the phase fluctuations. The other method is to measure the time-domain impulse response of the photodetector and to use a Fourier transform to calculate the phase difference as the power varies. The advantage of using a phase bridge is high precision, but the experimental setup is more complicated than measurement of the time domain impulse response. Measurement of the time domain impulse response, followed by the Fourier transform, is easier to do, but it is necessary to measure more average current points.

We will  simulate the impulse response of the photodetector to calculate the phase.  In the model, the input light is a pulse train. We calculate the output current as  a function of time. Then, we use a Fourier transform to calculate the RF phase in radians for the given Fourier frequency. We use the phase information to calculate the AM-to-PM conversion coefficient that is defined in Eq.~\ref{AMPM}. 

\begin{figure}[!h]
	\centerline{
		{\includegraphics[width=7.9cm]{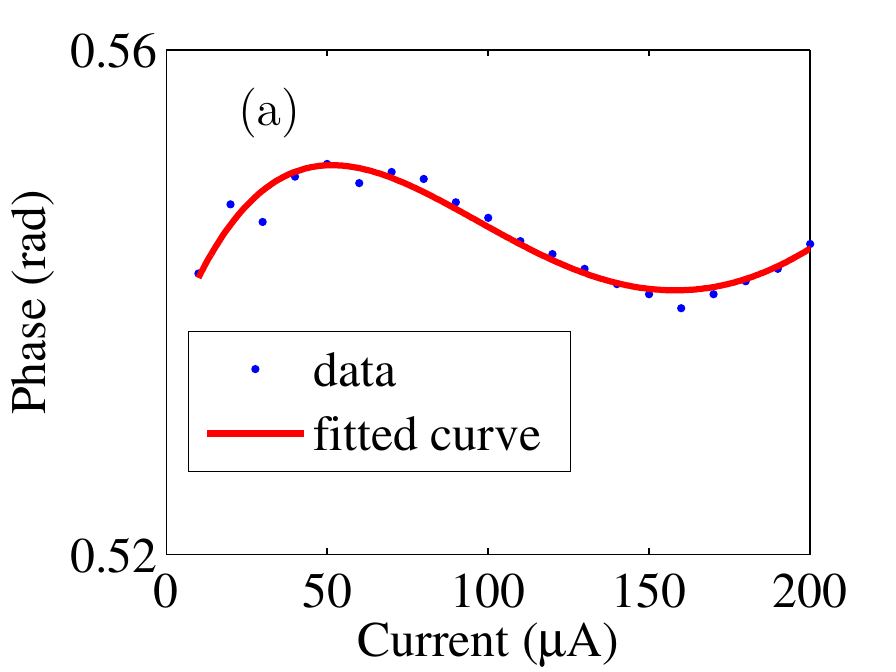}\label {phase-exp}}
		{\includegraphics[width=7.9cm]{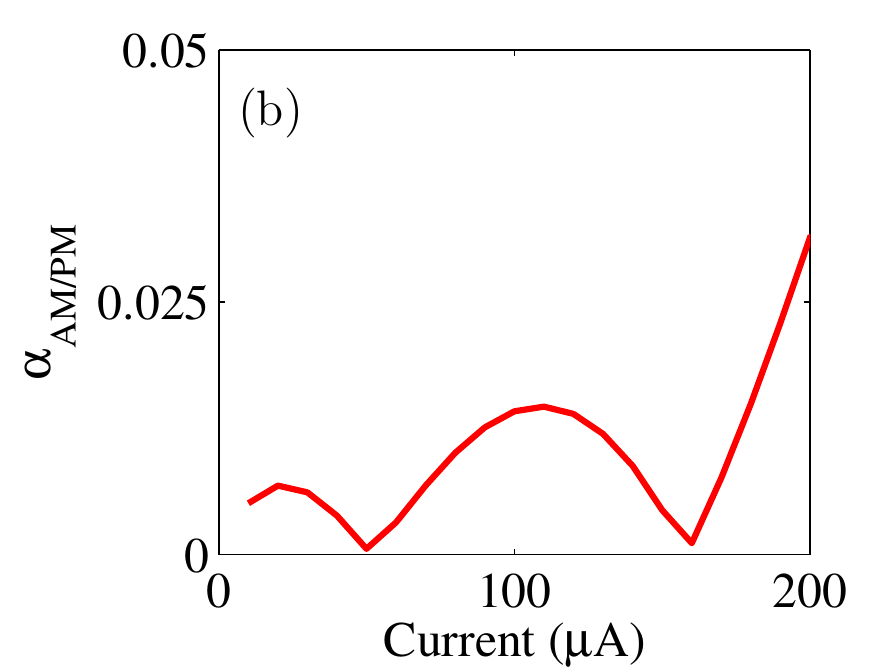}\label {phase-exp}}
	}
	\caption{The measured phase change and AM-to-PM noise conversion coefficient in the MUTC as a function of the output average photocurrent at 1 GHz. The bias voltage applied to the photodetector is 3 V.\label{exp-data} }
\end{figure}

In order to characterize phase change of MUTC photodiodes, time domain impulse responses of the photodiodes are measured based on a fiber pulse laser from Toptica Photonics with 100 femtosecond pulse width and a 50 GHz Agilent Infiniium digital sampling oscilloscope. A 25 dB RF attenuation was incorporated in the signal path since a large dynamic range was required in this measurement and the input of the oscilloscope was limited to 0.8 V maximum. A simple variable optical attenuator based on a precision blocking screw was used to tune the input optical power to photodiodes. Figure \ref{exp-data} shows the experimental phase change and AM-to-PM noise conversion coefficient.  The diameter of the device is 40 $\mu$m. The red curves are the fitted curves. The repetition rate is 250 MHz. We calculate the phase change and the AM-to-PM noise conversion coefficient at 1 GHz. We found that there are two nulls in the AM-to-PM noise conversion coefficient. The phase increases when the photocurrent increases to 50 $\mu$A, and next decreases when the photocurrent increases to 155 $\mu$A, and finally increases again when the photocurrent further increases. 

The pulse profile that we use  in our calculations is
\begin{equation}
y(t)=A\mathrm{sech}\left( \frac{t-t_p}{t_w}\right), \label{eq1}
\end{equation}
where $t_p$ is the pulse position, $t_w$ is the pulse duration, and $A$ is the pulse amplitude.

\section{Simulation results}
\begin{figure}[!h]
	\centerline{
		{\includegraphics[width=7.2cm]{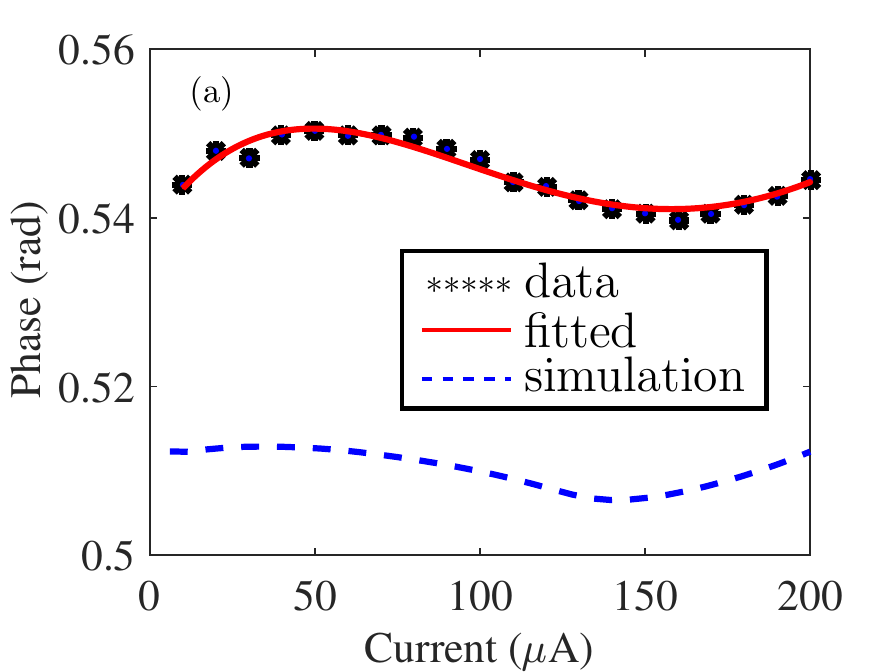}\label {phase1}}
		{\includegraphics[width=7.2cm]{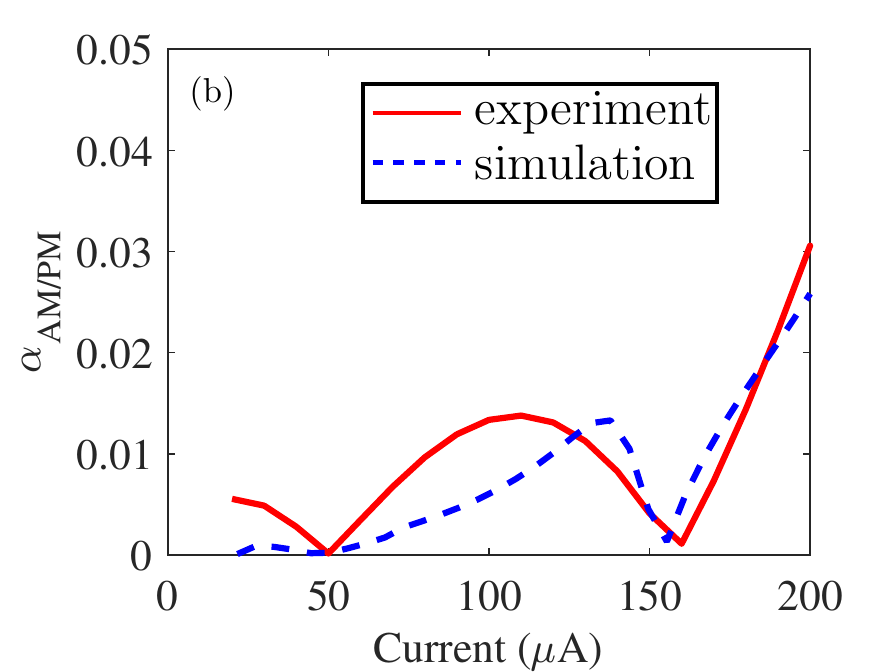}\label {amtopm1}}
	}
	\caption{The calculated (a) phase change and (b) AM-to-PM noise conversion coefficient in the MUTC device as a function of output average photocurrent at 1 GHz. The repetition rate is 250  MHz.\label{sim-data} }
\end{figure}

\begin{figure}[!h]
	\centerline{
		{\includegraphics[width=7.2cm]{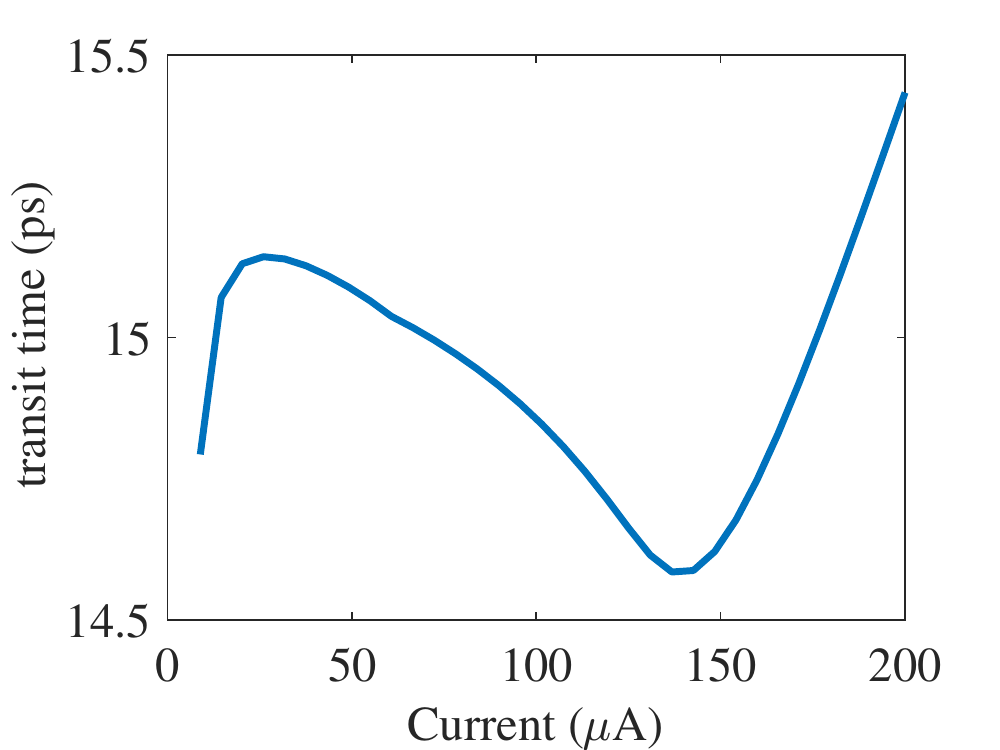}}
	}
	\caption{The calculated transit time in the MUTC device as a function of the average output  photocurrent. \label{transit} }
\end{figure}

\begin{figure}[!h]
	\centerline{
		{\includegraphics[width=7.2cm]{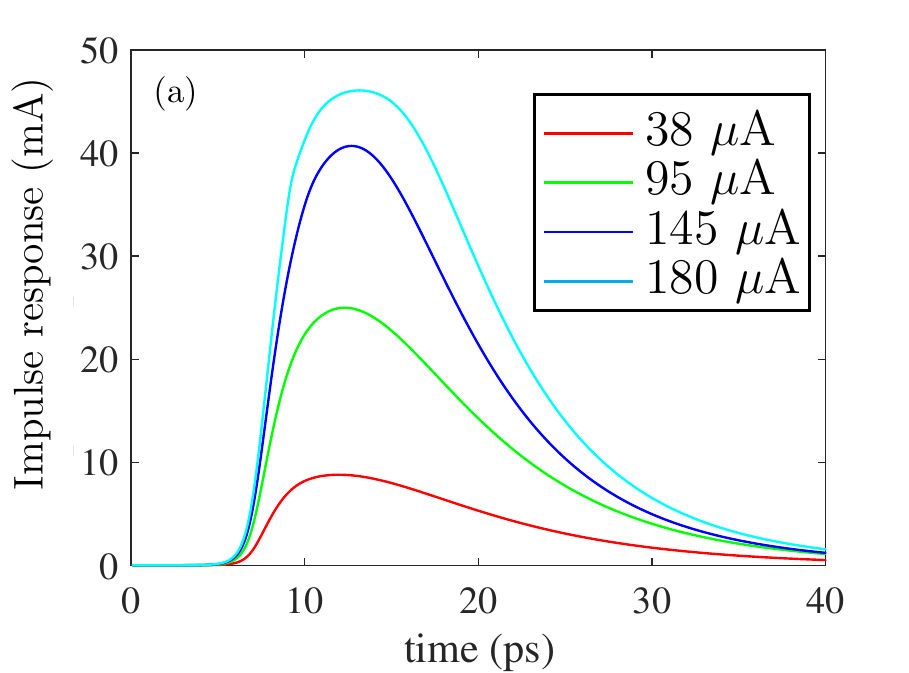}\label {response1}}
		{\includegraphics[width=7.2cm]{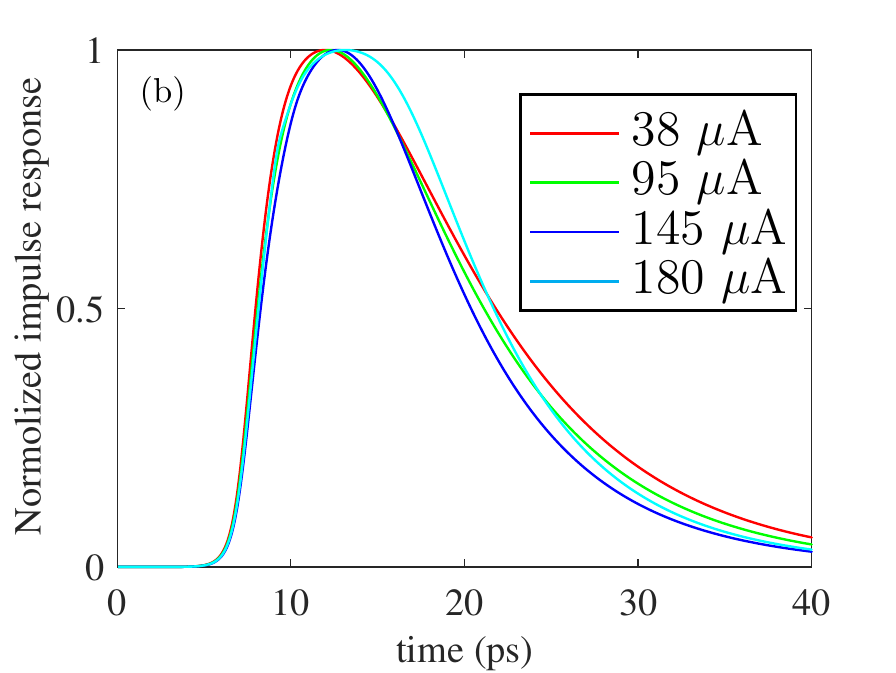}\label {response2}}
	}
	\caption{The calculated impulse response in the MUTC device for different average output photocurrents. \label{response} }
\end{figure}

\begin{figure}[!h]
	\centerline{
		{\includegraphics[width=7.2cm]{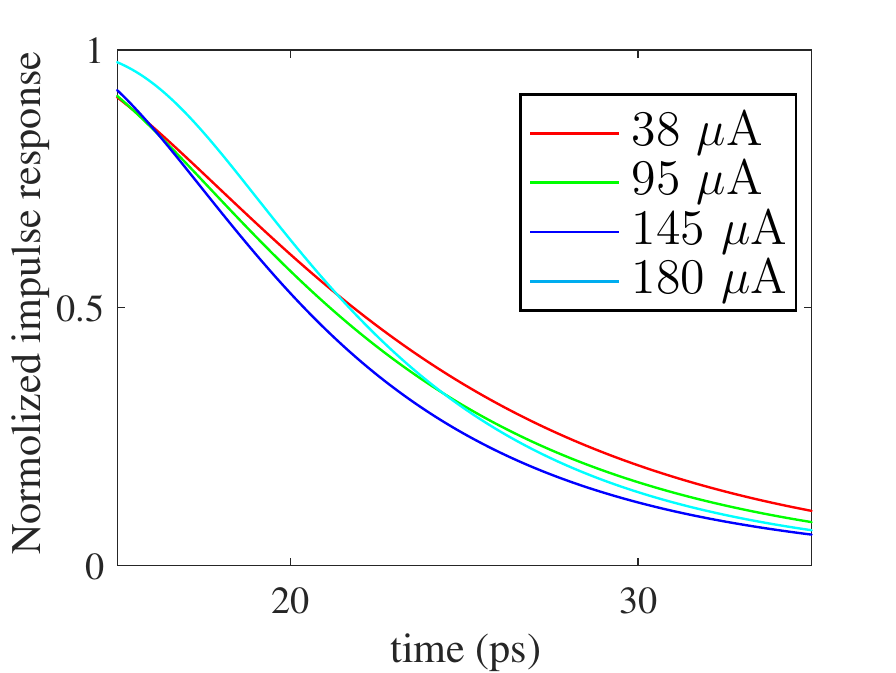}\label {response4}}
	}
	\caption{The calculated impulse response in the MUTC for different average output photocurrents. \label{response4} }
\end{figure}


Figure \ref{sim-data} shows the calculated phase change and AM-to-PM noise conversion coefficient. We obtained agreement with experiments for the phase change and AM-to-PM noise conversion coefficient. The nulls appear due to the phase changes when the photocurrent increases. We note that the offset in the phase between the simulation and the experimental data us due to the different input pulse phase. What we care about is the relative phase change as a function of average pulse current. The difference between measured phase and simulated phase may result from the RF cable, bias tee and RF attenuator which connect photodetector and oscilloscope; however this difference does not affect our conclusions.  Figure \ref{transit} shows the calculated average transit time in the device. We found that the behavior of the transit time is similar to the phase change. When the phase change increases, the transit time increases. At 38 $\mu$A, the transit time is a maximum. Figure \ref{response}(a) shows the impulse response for different average output photocurrents. Figure \ref{response}(b) shows the normalized  impulse response.  Figure \ref{response4} shows the tail of the normalized impulse response. When the average current increases, the time for the impulse response to reach its peak increases. However, when the current increases from 30 $\mu$A to 150 $\mu$A, the tail of the impulse response decreases. These two processes compete with each other, leading at first to a small phase increase and then a phase decrease. When the current is larger than 150 $\mu$A,  the output photocurrent response width increases dramatically, which corresponds to a phase increase. 

\section{Physical reason for the nulls}

The AM-to-PM noise conversion coefficient $\alpha$, defined by Eq.~(\ref{eq1}), depends on the phase change $\Delta \phi$. The phase change in turn depends on the average electron transit time in the device. As a major carrier in undepleted absorption region, holes will decay with a dielectric relaxation time which is much smaller than the electron transit time.

\begin{figure}[!h]
	\centerline{
		{\includegraphics[width=10cm]{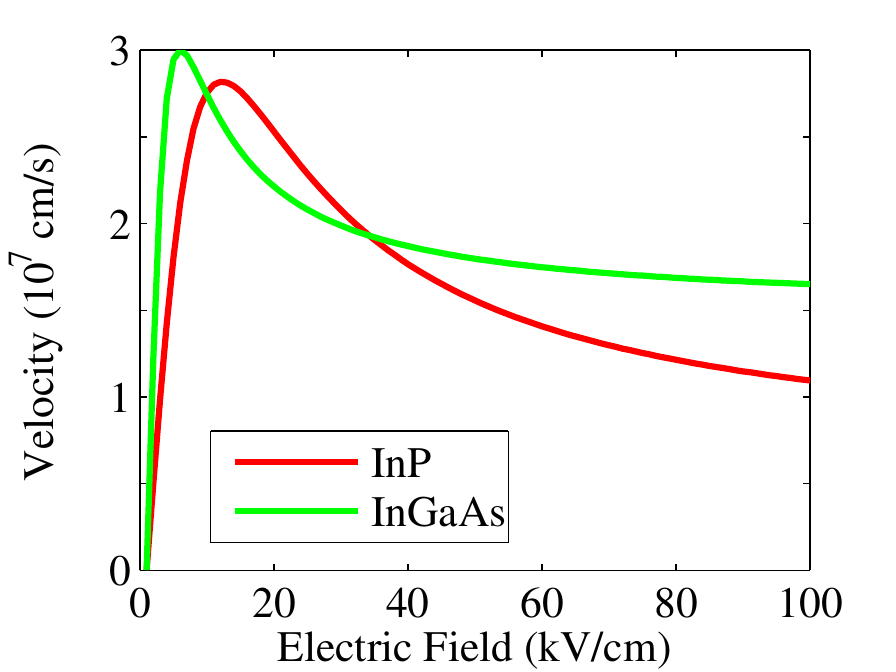}}
	}
	\caption{The calculated electron velocity as a function of electric field for InGaAs and InP. \label{Velocity1} }
\end{figure}

\begin{figure}[!h]
	\centerline{
		{\includegraphics[width=10cm]{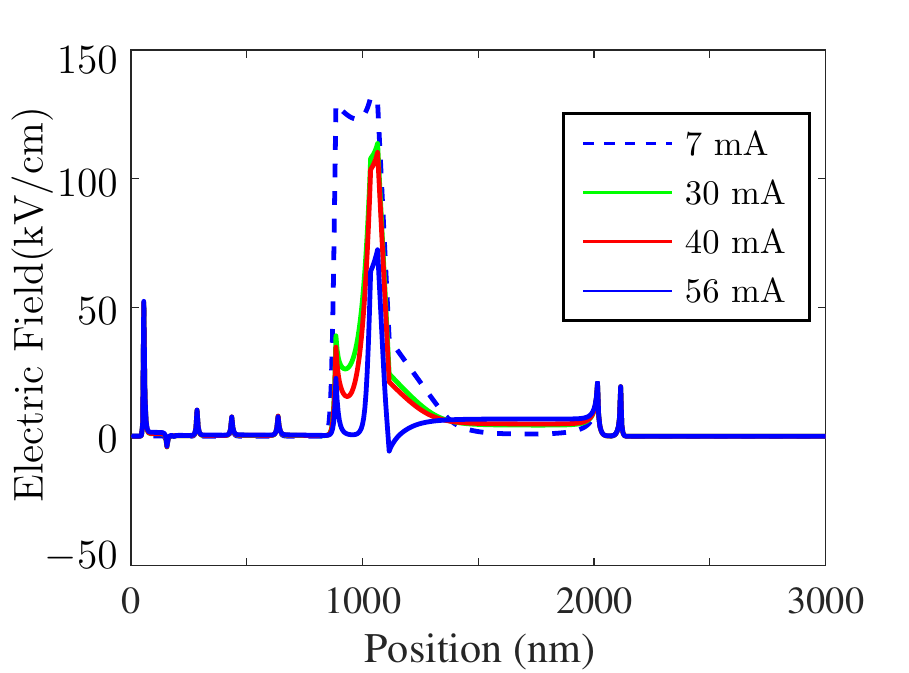}}
	}
	\caption{The calculated steady-state electric field distribution in the device for different photocurrents. The applied reverse bias is 3 V.  \label{Efield1} }
\end{figure}

\begin{figure}[!h]
	\centerline{
		{\includegraphics[width=7.2cm]{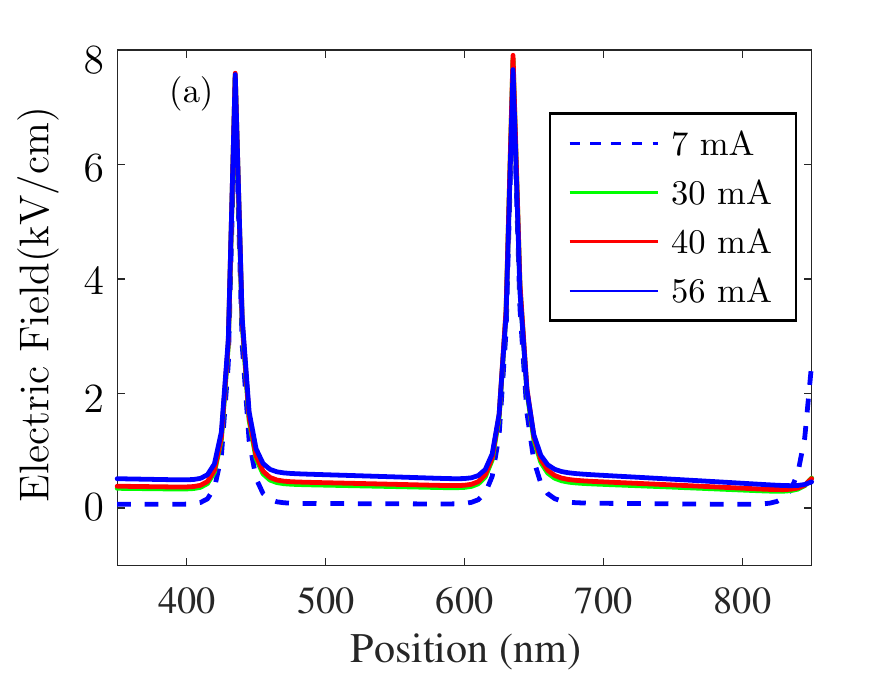}}
		{\includegraphics[width=7.2cm]{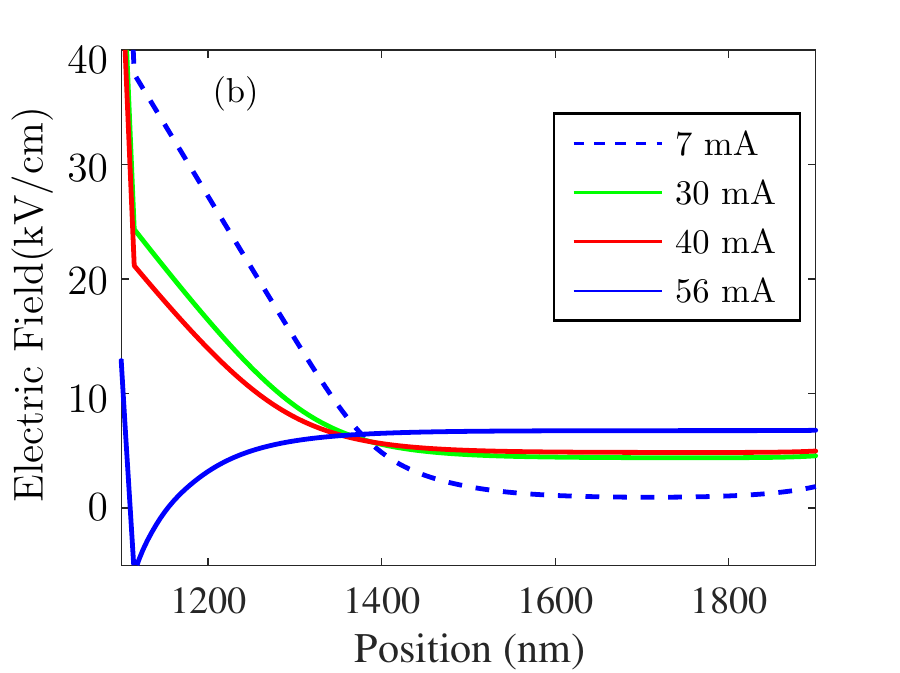}}
	}
	\caption{The calculated steady state electric field distribution in the device for different photocurrents in (a) the $p$-absorption region and (b) the collection region shown in Fig.~\ref{Structure}. The applied reverse bias is 3 V.  \label{Efield2} }
\end{figure}

Figure \ref{Velocity1} shows the velocity as a function of electric field for InGaAs and InP.  The velocity first rapidly increases and then decreases as a function of the electric field, eventually saturating. The principal physical reason for this behavior is that the ratio of heavy ($X$- and $L$-valley) electrons to light ($\Gamma$-valley) electrons increases as the electric field increases, slowing the average velocity. This behavior is important in understanding the appearance of the higher-current null in the AM-to-PM conversion.

There is an intrinsic absorption region and a $p$-absorption region in the device. The photogenerated electrons in the intrinsic region travel fast. However, there is a heterojunction between InGaAs and InP. When the pulse energy increases, more electrons accumulate at the heterojunction. So, a larger optical pulse energy leads to a large time for the photocurrent to reach its peak.

The output photocurrent decay time changes significantly when the optical pulse energy changes because the peak photocurrent changes, leading to a change in the electric field in the device. The velocity depends on the electric field and changes as the peak current varies. When the average photocurrent is 38 $\mu$A and 145 $\mu$A, the corresponding peak currents are 10 mA and 40 mA. Figure~\ref{Efield1} shows the electric field in the device for different photocurrents. Figure~\ref{Efield2}(a) shows the electric field in the $p$-absorption region, and collection region shown in Fig.~\ref{Structure}.  We observe that when the photocurrent increases in the device, the electric field in the $p$-absorption region and in the collection region increases. Figure~\ref{Velocity1} shows the velocity as a function of the electric field for InGaAs and InP.  The electron velocity increases in the $p$-absorption region. The decay times decrease as the output photocurrent increases. The two processes compete with each other. When the output average photocurrent is less than 38 $\mu$A, the time that it takes the photocurrent to reach its peak dominates the total transit time. So, the phase increases as the photocurrent increases up to 38 $\mu$A. When the output photocurrent is between 38 $\mu$A and 145 $\mu$A, the decay time dominates the average transit time. Hence, the phase increases when the photocurrent increases. Finally, when the output photocurrent is larger than 145 $\mu$A, which corresponds to a peak current of 40 mA, the electric field in the $n$-region, which shows in Fig.~\ref{Efield2}(b), becomes negative, so that the electron transit time increases when the output photocurrent increases. Hence, the phase again increases when the photocurrent increases.

\subsection{Pulse duration}
\begin{figure}[!h]
	\centerline{
		{\includegraphics[width=10cm]{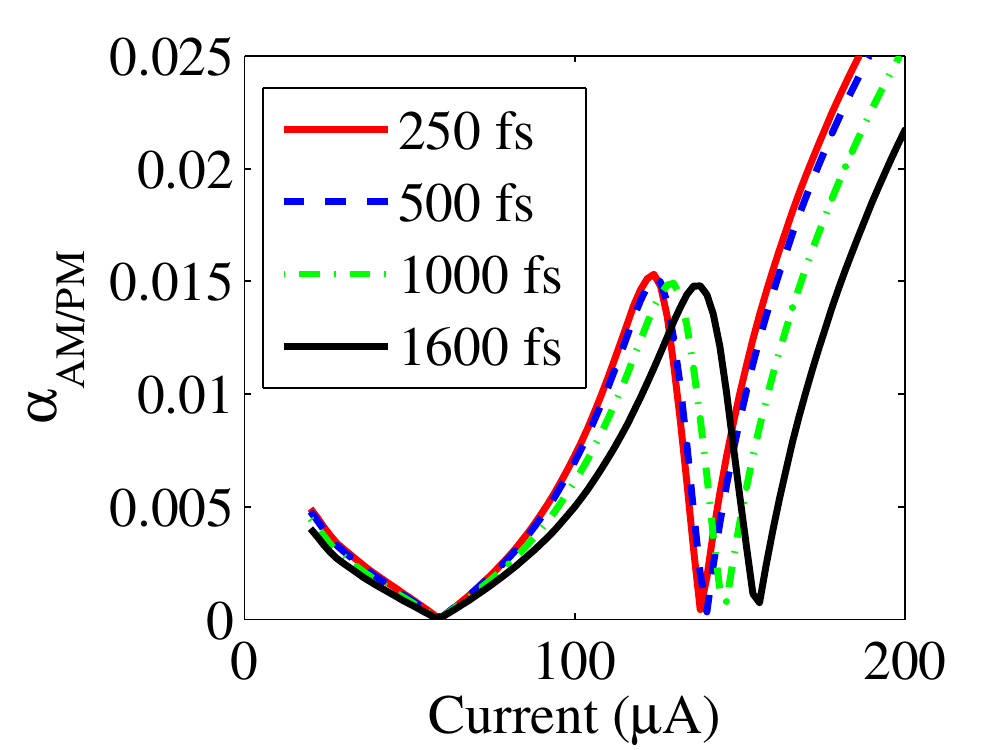}}
	}
	\caption{The calculated AM-to-PM noise conversion with different pulse durations at 1 GHz. \label{PulseWidth} }
\end{figure}

\begin{figure}[!h]
	\centerline{
		{\includegraphics[width=10cm]{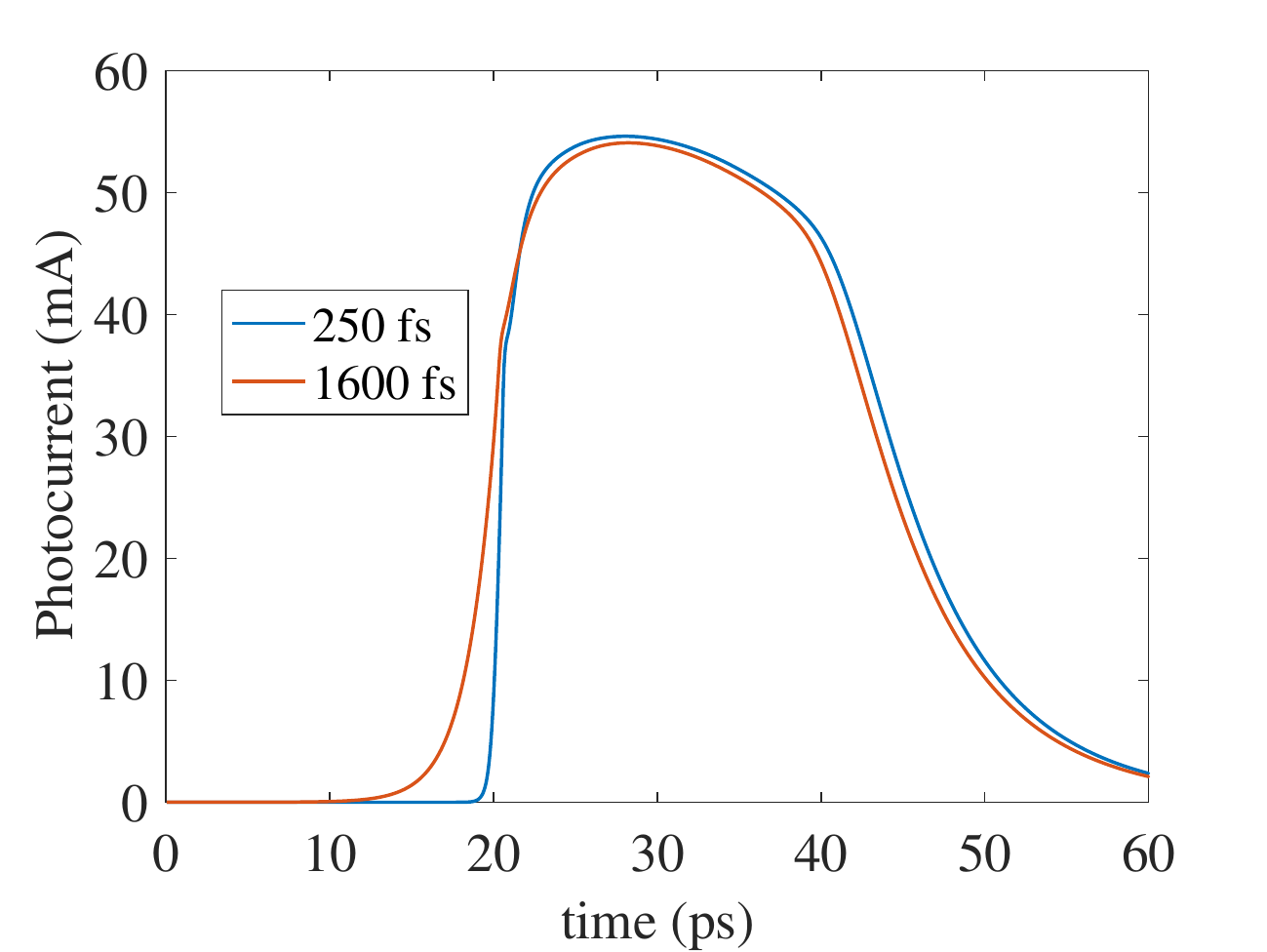}}
	}
	\caption{The calculated impulse response for different pulse durations and the same pulse energy. \label{response3} }
\end{figure}

Figure~\ref{PulseWidth} shows the AM-to-PM noise conversion coefficient as a function of average photocurrent for different pulse durations. We observe that the null position does not change when the pulse width is less than 500 fs. When the pulse duration is larger than 500 fs, even if it is much smaller than the photodetector response time, which is 40 ps for this photodetector, the second null position shifts to larger photocurrents. When the pulse duration increases, the peak current decreases. Figure \ref{response3} shows the impulse response for different pulse durations with the same pulse energy. The photocurrent at which the second null occurs depends on the electric field,  which is determined by the peak current. When the pulse duration increases, the peak current decreases, and a second null appears at a larger average photocurrent.

\subsection{Repetition rate}

\begin{figure}[!h]
	\centerline{
		{\includegraphics[width=8cm]{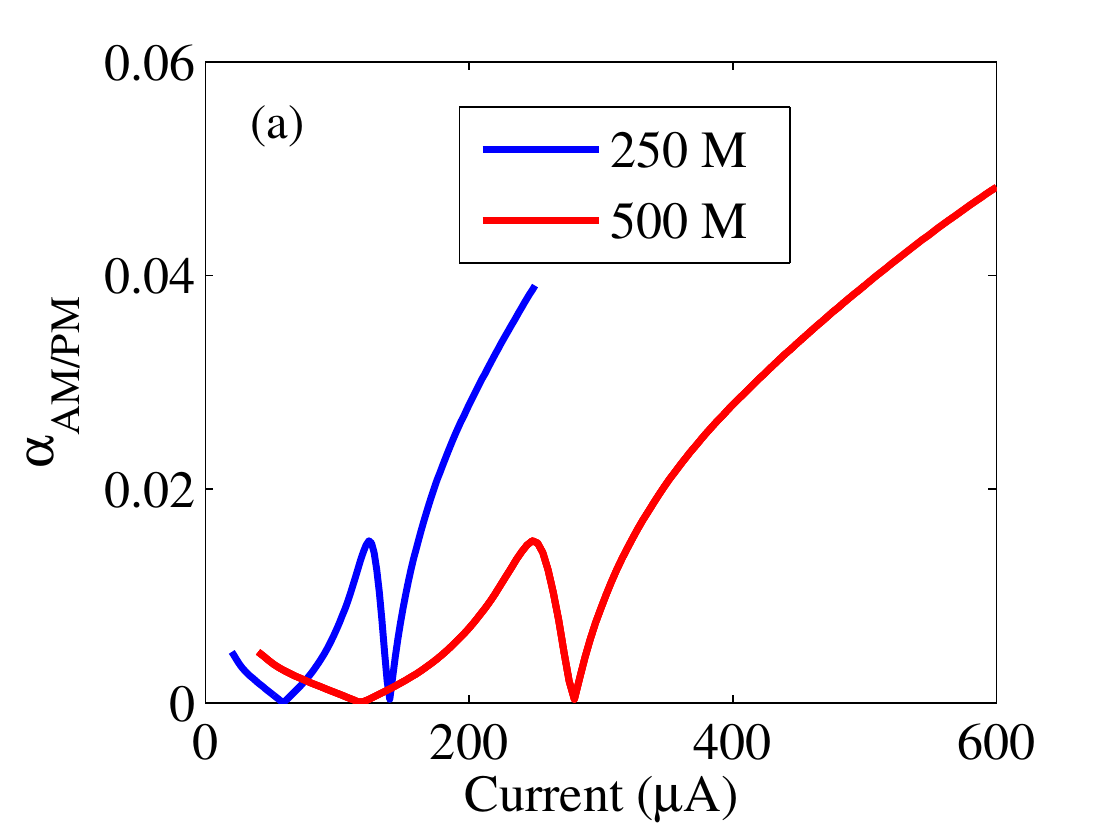}}
		{\includegraphics[width=8cm]{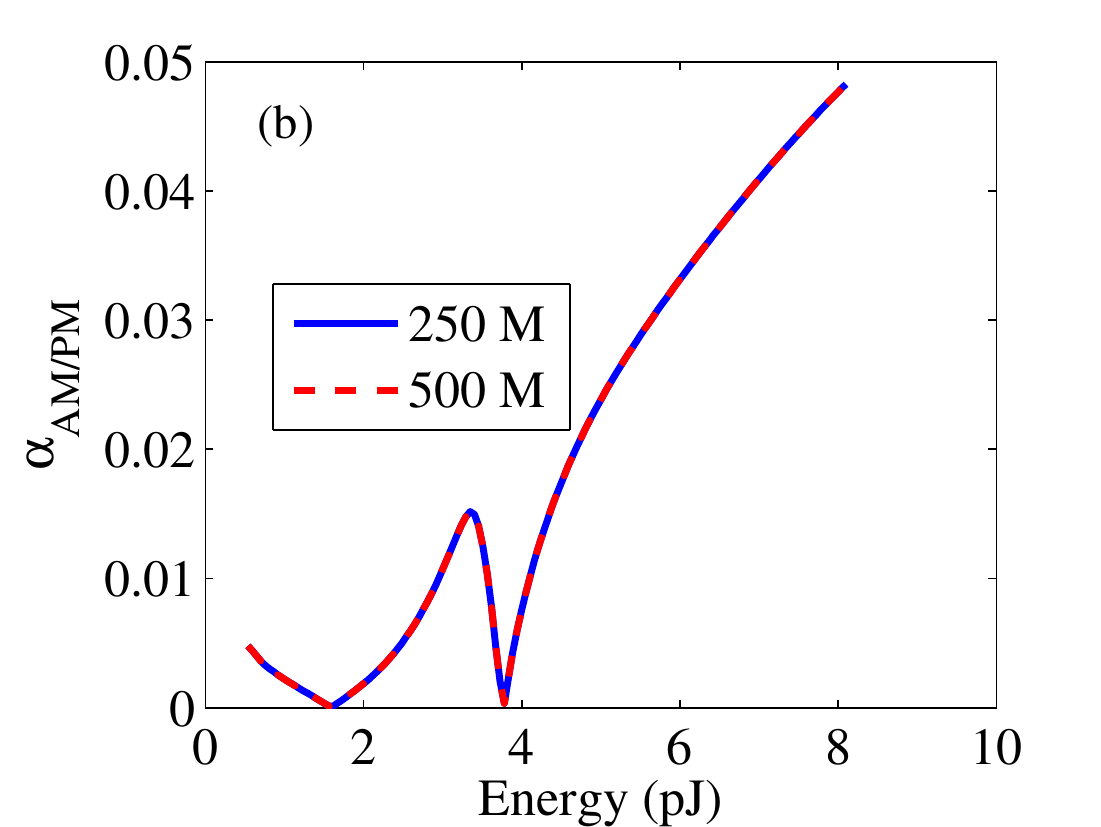}}
	}
	\caption{The calculated AM-to-PM noise conversion coefficient at 1 GHz in the MUTC for different repetition rate. (a) The coefficient as a function of average current. (b) The coefficient as a function of energy. \label{ampm-repe} }
\end{figure}

Figure \ref{ampm-repe}(a) shows the AM-to-PM noise conversion coefficient as a function of average current for different repetition rates. The curves are the same except for a scale factor, as shown in Fig.~\ref{ampm-repe}(b), where we plot the two curves as a function of optical pulse energy. So, the AM-to-PM noise conversion coefficient does not depend on the repetition rate when the repetition rate is less than the bandwidth of the photodetector, which is about 25 GHz. The average photocurrent depends on the repetition rate. This result is consistent with experiments \cite{Fortier2013}.

\begin{figure}[!h]
	\centerline{
		{\includegraphics[width=10cm]{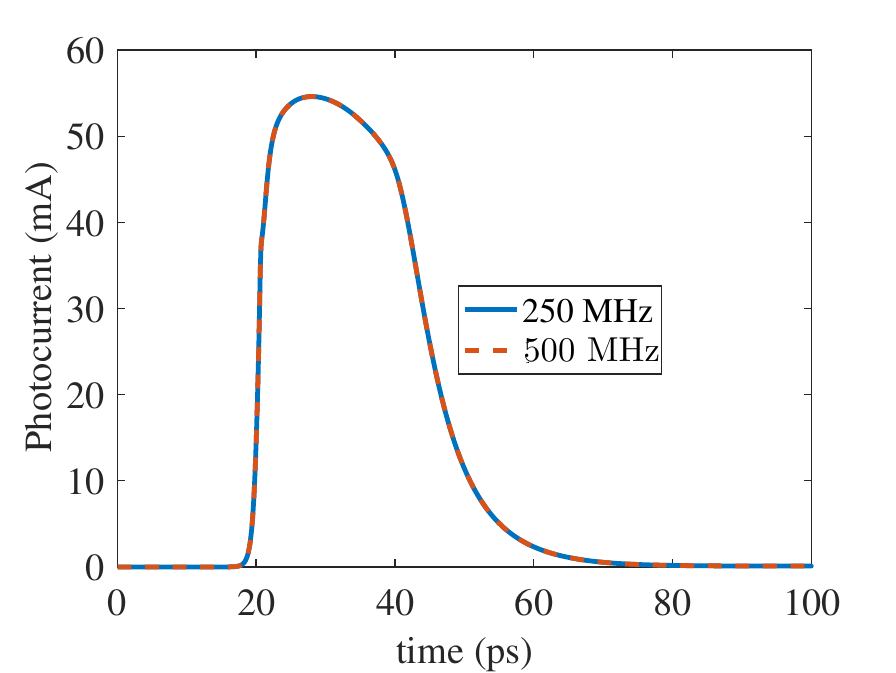}}
	}
	\caption{The calculated impulse response of the MUTC device for different repetition rates. \label{response-rate} }
\end{figure}

Figure \ref{response-rate} shows the impulse response for different repetition rates. We observe that the shape of the output photocurrents are the same. The only difference is that the output photocurrent pulses are spaced twice as far apart, lowering the average photocurrent by a factor of two, which has no effect on the Fourier transform of the photocurrent impulse response.
\section{Discussion}
\subsection{Suggestions for improvement}
\begin{figure}[!h]
	\centerline{
		{\includegraphics[width=7.2cm]{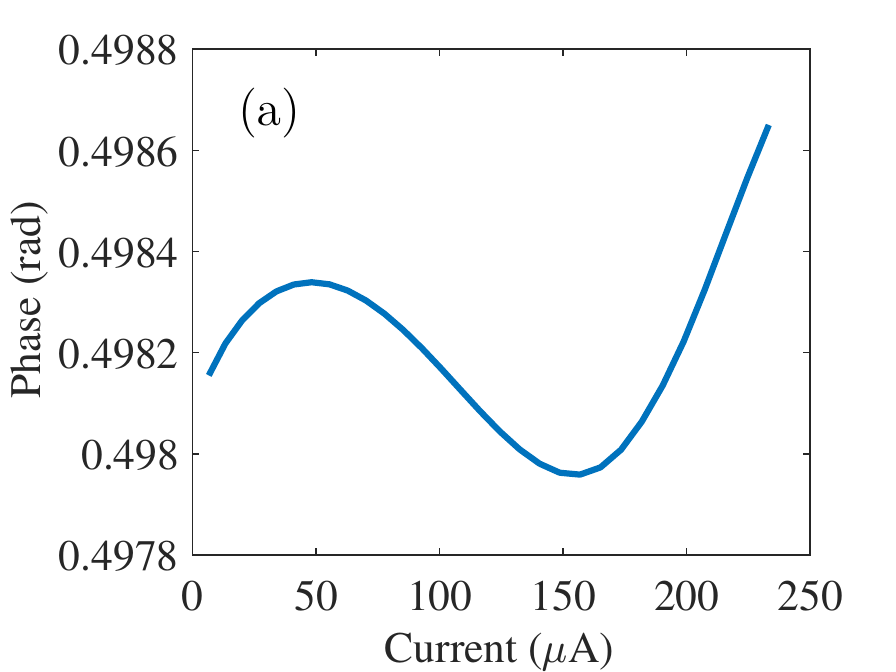}\label {phasebf0}}
		{\includegraphics[width=7.2cm]{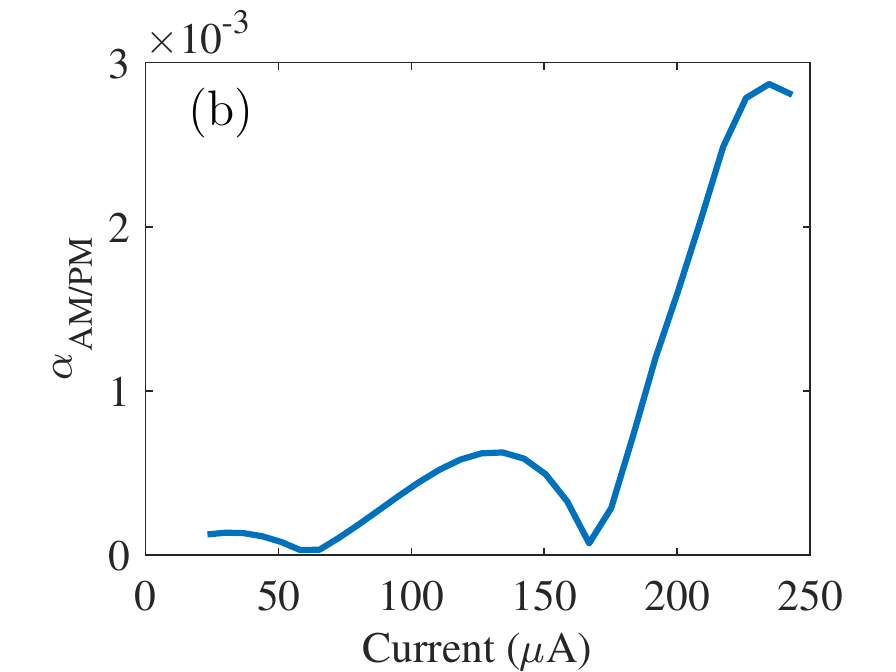}\label {amtopmbf0}}
	}
	\caption{The calculated (a) phase change and (b) AM-to-PM noise conversion coefficient in the MUTC device as a function of output average photocurrent at 1 GHz. The heterojunction effect is not included in the model. \label{resultbf0} }
\end{figure}

The AM-to-PM noise conversion coefficient depends on the phase change in the device, which is due to the change in the average electron transit time. The pileup of electrons that occurs at the heterojunction between the intrinsic absorption layer and the cliff layer is a major factor increasing the transit time. It is not possible in practice to completely to eliminate the heterojunction. However, by having several intermediate layers, it should be possible to improve the heterojunction transition. Figure~\ref{resultbf0} shows the phase change and AM-to-PM conversion coefficient if there is no heterojunction between InGaAs and InP. We can see that $\alpha_{AM/PM}$ is reduced by about one order of magnitude. 

\section{Conclusion}
We used a drift-diffusion model to study the AM-to-PM conversion in an MUTC photodetector. There are two nulls in the AM-to-PM as a function average photocurrent. We have explained the appearance of these nulls as a consequence of changes in the transit time through the device, which is in turn due to the nonlinear dependence of the electron velocity on the electric field.

We also show that when the pulse duration is less than 500 fs, the AM-to-PM conversion coefficient does not change. When the pulse duration is greater than 500 fs, the second null in the AM-to-PM shifts to larger photocurrents. The repetition rate does not change the AM-to-PM conversion coefficient when plotted as a function of the input optical pulse energy.

The AM-to-PM noise conversion coefficient can be greatly reduced by having an intermediate transition layer between the heterojunction InGaAs and InP. 

\section*{Acknowledgements}
The authors wish to thank the anonymous reviewers for their valuable suggestions. Work at UMBC was partially supported by the Naval Research Laboratory.  The simulations were carried out at UMBC's high performance computing facility.




\end{document}